\documentclass[aps,pra,nofootinbib,superscriptaddress,showpacs,reprint,floatfix]{revtex4-2}

\usepackage{sidecap}
\usepackage{graphicx}
\usepackage{amsmath}
\usepackage{float}
\usepackage{booktabs}
\usepackage{threeparttable} 
\usepackage{amssymb}
\usepackage{longtable}
\usepackage{adjustbox}
\usepackage{diagbox}
\allowdisplaybreaks
\pagebreak

\usepackage{makeidx}
\usepackage{amsfonts}
\usepackage[usenames,dvipsnames]{pstricks}
\usepackage{subfigure}
\usepackage{epstopdf}
\usepackage{float}
\usepackage{booktabs}
\usepackage{threeparttable} 
\usepackage{pst-grad} 
\usepackage{mathtools}
\usepackage[colorlinks,hyperindex]{hyperref}
\usepackage{array}
\usepackage{longtable}

\begin{document}
\title{Enhancement of cell membrane poration by the antimicrobial peptide Melp5}
\author{Qixuan Li\textsuperscript{1}} 
\author{Xiaoshuang Zhong\textsuperscript{1}} 
\author{Liang Sun\textsuperscript{*}} 
\author{Liang Dai\textsuperscript{*}}

\pagestyle{empty}

\begin{abstract}
Melittin, a natural antimicrobial peptide comprising 26 amino acid residues, can kill bacteria by inducing pores in cell membranes. Clinical applications of melittin as an antibiotic require a thorough understanding of its poration mechanism and mutations that enhance its antimicrobial activity. Previous experiments showed Melp5, a variant of melittin with five mutations, exhibits a higher poration ability. However, the mechanism of the enhanced poration ability is not fully understood. Here, we investigated the mechanism by comparing the poration of melittin and Melp5 using coarse-grained (CG) and all-atom (AA) molecular dynamics (MD) simulations. We observe that Melp5 is likely to form a pore with 5 peptides (pentameric), while melittin is likely to form a pore with 4 peptides (tetrameric). Our atomistic MD simulations show that the pentameric pore of Melp5 has a higher water permeability than the tetrameric pore of melittin. We also analyze the stability of the pores of melittin and Melp5 by calculating the interaction energies of the pores. In particular, we investigate the effects of mutant residues on pore stability by calculating electrostatic and LJ interactions. These results should provide insights on the enhanced poration ability of Melp5 and push it toward applications.
\end{abstract}

\hspace*{\fill}

\maketitle
\section{Introduction}
Antimicrobial peptides (AMPs) have garnered significant interest as a potentially effective group of therapeutic agents in the fight against antimicrobial resistance. Recent research has revealed that AMPs can exert antimicrobial properties by compromising microbial cell membrane structural integrity. This disruption occurs through various mechanisms, including pore formation, membrane lysis, and interference with crucial cellular processes. Consequently, these actions result in the eradication of the specific bacteria being targeted \cite{1,2,3,4,5,6}.

Melittin, an amphipathic peptide with a $\alpha$-helix structure, has been extensively studied and is known to spontaneously create openings in bacterial cell membranes, resulting in cytoplasm leakage. This phenomenon has garnered significant attention and has been the focus of extensive research \cite{7,8,9,10,11}. Similar to other AMPs, melittin is composed of 26 amino acid residues. The N-terminal portion of the peptide is hydrophobic, while the C-terminal portion is polar and positively charged. This structural arrangement results in the formation of the peptide head and tail, as depicted in Fig.\ \ref{fig1}(a) and (b). Several previous experiments \cite{12,13,14,15} and computer simulations \cite{16, 17, 18, 19} have provided evidence supporting the hypothesis that the antimicrobial properties of melittin can be attributed to its ability to induce the formation of transmembrane pores, facilitate aggregation, dissolve membranes, and extract lipids from the membrane. Despite melittin's considerable potential as an antimicrobial peptide, it also has several drawbacks and limitations. Their limitations include potent cytotoxic effects \cite{20} and limited penetration capacity \cite{21}, limiting its therapeutic potential. Furthermore, melittin's clinical application is restricted due to concerns surrounding its stability and immunogenicity. To tackle these concerns, Melp5, a synthetic derivative that has undergone modifications based on melittin, has been developed \cite{22,23,24,25,26}. Melp5 has amino acid residue mutations at five specific sites, namely ALA-10, ALA-22, ALA-23, GLN-24, and LEU-26. Woo and Lee \cite{27} demonstrate that Melp5 exhibits an enhanced ability to insert into the membrane and produce larger transmembrane pores in comparison to melittin. This superiority is observed even when considering very low peptide concentrations and peptide-to-lipid ratios.

Through a comparative analysis of Melp5 and melittin amino acid residues, it becomes evident that Melp5 exhibits a higher count of hydrophobic amino acid residues (20 versus 16). Melittin has higher electrostatic charges than Melp5, with melittin having +6 charges and Melp5 having +3 charges. As a result, Melp5 exhibits enhanced hydrophobicity and diminished positive charge. This leads to an augmentation of the van der Waals interaction with the lipid bilayer and a reduction in electrostatic repulsion, respectively. The impact of amino acid sequence substitutions on membrane activity is crucial to enhancing Melp5's antimicrobial ability as a transmembrane peptide \cite{22,27}.

Molecular dynamics (MD) simulations have proven to be highly valuable in the study of interactions between AMPs and lipid bilayers \cite{28,29}. In recent years, Yuan et al.\ \cite{30} have employed MD simulations to demonstrate that the introduction of a melittin monomer penetrating into a transmembrane pore containing three or more peptides in the lipid bilayer results in the formation of a water channel with an increased radius. Their work underscored the significance of accurately quantifying free energy barriers to determine the rates at which AMPs form pores. However, the utilization of MD simulations to investigate these interactions poses a significant challenge. The process of AMPs spontaneously penetrating the lipid bilayer requires a long simulation period, often exceeding 1000 ns. As a result, all-atom MD simulations become computationally challenging.

Nowadays, coarse-grained (CG) MD simulations have emerged as a valuable tool for simulation efficiency. In recent years, CG force fields have been developed specifically for the investigation of peptides and lipid bilayers \cite{31,32,33,34}. The Martini force field is widely utilized and has demonstrated considerable success in various applications. Santo et al.\ \cite{35,36} performed a variety of Martini CG MD simulations to examine the penetration mechanisms of two AMPs, namely magainin-2 and melittin, within phosphatidylcholine bilayers. In addition, Liu and colleagues \cite{19} conducted a study that integrated experimental techniques with Martini CG MD simulations. Their research aimed to elucidate the intermediate states that occur during pore formation on a cell membrane induced by melittin. Furthermore, Sun et al.\ \cite{37,38} employed the Martini force field to investigate the interactions between melittin and lipid bilayers. Their results revealed the importance of a precursor defect in the membrane and a cooperative mechanism of melittin in stabilizing the pores. The application of the Martini force field to investigate the AMP-bilayer system has produced successful results \cite{39,40,41}. Nevertheless, a drawback of CG simulations is the exclusion of atomistic intricacies. In order to overcome this constraint, it is possible to employ a technique known as backmapping, wherein the CG structures derived from CG MD simulations are converted into all-atom structures, which are then used for all-atom MD simulations.

In this work, we utilize CG MD simulations employing the Martini 3.0 force field to investigate the spontaneous aggregation of melittin and Melp5 in the membrane. Our work suggests that Melp5 exhibits a higher propensity for the formation of oligomers, particularly pentameric pores, while melittin is more inclined to form tetrameric pores. Through calculating the residue-membrane LJ interactions and comparing the sustained stable peptide aggregation time of specific oligomers, we find that the mutant residues of Melp5 are indispensable for enhanced pore stability. To further examine the enhanced poration ability exhibited by Melp5, we compare the water permeability and pore radius between pentameric pores of Melp5 and tetrameric pores of melittin in all-atom simulations with CHARMM36MM force field.

\section{Methods}
\textbf{CG MD simulation} 
\vspace*{0.5\baselineskip}

CG MD simulations are conducted using the GROMACS package \cite{42}, and the Martini force field (version 3.0) is employed \cite{43}. The lipid bilayer's initial structure is generated through the use of the CHARMM-GUI online server \cite{44}. The lipid bilayer is composed of two leaflets, with each leaflet containing 298 DLPC lipids. All DLPC lipids are asymmetrically distributed between two leaflets. The peptide aggregation in DLPC lipids has been observed in previous study \cite{27}. Melittin crystal structure is acquired from the PDB (PDB: 2MLT) and subsequently transformed into a CG melittin model. In order to determine the secondary structure of each residue in the Martini force field, the Dictionary of Secondary Structure of Proteins (DSSP) algorithm \cite{45} is employed. This algorithm identifies that the melittin peptide consists of two $alpha$-helixes connected by a kink region, specifically regarding the residue THR-11 and GLY-12 as the kink region.

In our simulations, a quantity of peptides, ranging from 1 to 25, is inserted vertically and randomly into the lipid bilayer (Fig.\ {\ref{fig1}(c) and (d)). In order to obtain statistically significant outcomes, we conduct 20 replicated simulations using distinct random seeds. This approach allows us to enhance the sampling of both aggregation and pore formation events Table\ \ref{tbl:table1}. The system is neutralized by the addition of ions (Na$^{+}$, CL$^{-}$).

Each simulation starts with energy minimization using the steepest descent method \cite{46}, followed by a series of equilibration NPT simulations. Harmonic position restraints are applied to the melittin and Melp5 backbones during NPT ensemble equilibration. These restraints are gradually weakened in five separate NPT simulations, resulting in a cumulative simulation time of 4.75 ns. Using the Berendsen coupling method \cite{47} with a coupling constant of 5 ps, the pressure is maintained at 1 bar. In order to achieve independent control over the pressure in the x-y directions and the z-direction, a semisotropic coupling scheme is applied. The V-rescale thermostat \cite{47} is employed to couple the temperature, with a reference temperature of 310 K and a time constant of 1.0 ps. Electrostatic interactions are estimated with the reaction-field method and default Martini parameters, with LJ interactions truncated at 1.2 nm. The pressure coupling method is transitioned to a Parrinello-Rahman barostat in the concluding stage of production \cite{48}. The time step is set at 20 fs, and each simulation iteration is conducted over 5000 ns. The generation of all simulation snapshots is accomplished through visual MD (VMD) simulations \cite{49}.

\begin{table*}[htbp]
\centering
\caption{Initial parameter setting of this work.}
\setlength{\tabcolsep}{5mm}{
\begin{tabular}{cccccccc} 
\toprule
\midrule
  \multicolumn{1}{c}{System} &   \multicolumn{1}{c}{Melittin} &  \multicolumn{1}{c}{Melp5} &  \multicolumn{1}{c} {Lipid bilayer} &  \multicolumn{1}{c}{Force field} &  \multicolumn{1}{c}{Replicas} &  \multicolumn{1}{c}{Simulation time}\\
  \midrule
  CG &  25 & 25 & DLPC & Martini 3.0 & 20 & 5000 ns \\
  AA &  4 & 5 & DPPC/POPG & CHARMM36M & 5 & 300 ns \\
  \midrule
  \bottomrule
  \end{tabular}}
  \label{tbl:table1}
\end{table*}

\vspace*{0.5\baselineskip}
\textbf{All-atom MD simulations}
\vspace*{0.5\baselineskip}

In order to evaluate the pore radius and water permeability of oligomers obtained from CG simulations, we utilize all-atom MD simulations. The initial step involves the use of a backmapping procedure to convert the final structures obtained from CG structures into atomistic structures. These atomistic structures are used as initial configurations for all-atom simulations. In order to replicate the cellular environment, a lipid bilayer composed of 216 DPPC and 88 POPG lipids is introduced into each leaflet for all-atom MD simulations. Thus, depositing oligomers into this kind of lipid bilayer can precisely capture the formation of transmembrane pores in the cell membrane.

Here, we utilize the GROMACS package and the CHARMM36M force field for all-atom simulations \cite{50}. The particle mesh Ewald method is implemented to handle electrostatic interactions in our work \cite{51}. Additionally, a cutoff distance of 1.2 nm is utilized to account for nonbonded electrostatic interactions. In order to ensure a consistent 325 K temperature, the Nose-Hoover coupling scheme is implemented. Additionally, a Parrinello-Rahman barostat is applied \cite{48}, to maintain a pressure of 1.0 bar through isotropic coupling. In order to maintain numerical stability, the LINCS algorithm constrains all hydrogen atom bonds \cite{52}. In the course of the simulations, a temporal increment of 2 fs is employed, and the duration of each simulation is 300 ns.

\vspace*{0.5\baselineskip}
\textbf{Radial distribution function}
\vspace*{0.5\baselineskip}

In the microscopic structure, radial distribution function (RDF) is widely used to describe the local density of specific atomic groups $a$ at a distance $r$ from specific atomic groups $b$. During the final 10 ns of CG simulations, various stable or metastable oligomers have been observed with our eyes. Oligomers (hexamers and heptamers) always stay in the metastable state, which makes their existence time much lower than those of trimers, tetramers, and pentamers.  To quantify the time uniformly, we use MD trajectories that could observe the oligomeric structure with stable peptide organization conformation at the final 10 ns of 100 frames. We choose the center of mass (COM) of all residues within the oligomers as atomic groups of $a$ and $b$ to calculate the residue-residue RDF as $g(r)$ by GROMACS.

\vspace*{0.5\baselineskip}
\textbf{The inner pore radius for water permeability}
\vspace*{0.5\baselineskip}

The determination of the radius of an aqueous pore involves quantifying the quantity of water molecules situated within the central region of a lipid bilayer, specifically within z coordinates ranging from -1 nm to 1 nm. The volume occupied by the water molecules is approximated as a cylindrical shape, and it is assumed that the density of water molecules within this cylinder is equivalent to that of bulk water. The estimation of the inner pore radius is accomplished through the utilization of the equation developed by Sun et al.\ \cite{17,53}:

\begin{equation}
\begin{aligned}
&r_{pore} =2\sqrt{\frac{18N_{w} }{602\pi H} }  \\
\end{aligned}
\end{equation}

Here, the variable representing the average number of water molecules in the cylinder over a given time period is denoted as $N_{w}$. The variable $H$ represents the height of the cylinder, while $r_{pore}$ denotes the radius of the aqueous pore.

\section{Results and Discussion}

\textbf{Spontaneous aggregation among the peptides}
\vspace*{0.5\baselineskip}

\begin{figure*}
\centering
\includegraphics[width=0.9\textwidth]{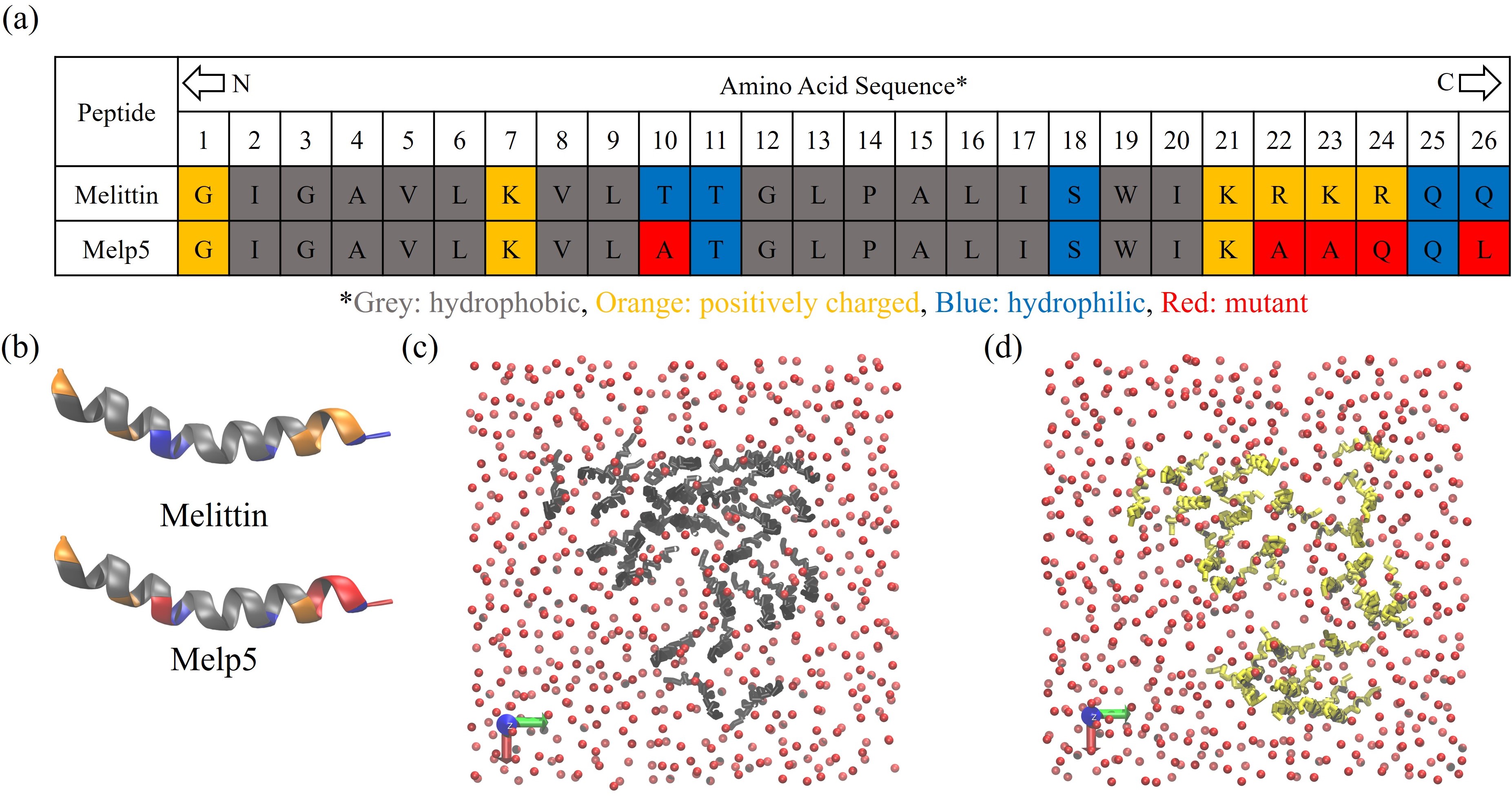}
\caption{
(a) The amino acid sequence of melittin and its synthetic derivative, Melp5. The utilization of distinct colors to represent residues of varying properties is noticed: grey to denote hydrophobicity, orange to signify positive charge, and blue to indicate hydrophilicity. The red residues observed in Melp5 correspond to the mutant sites derived from melittin. (b) The atomistic conformation of melittin and Melp5. Residues exhibiting distinct properties are visually distinguished through the application of corresponding colors in (a). (c) and (d) The initial frame of top view snapshot depicts the melittin (25 peptides) and Melp5 (25 peptides) are randomly and perpendicularly inserted into the lipid bilayers. The lipid bilayers are depicted with red balls representing the head groups, while the lipid tail groups are omitted from the representation. Melittin peptides are rendered as black and Melp5 peptides are rendered as yellow.}
\label{fig1} 
\end{figure*}

During CG MD simulations, a constant lipid profile is maintained, comprising 596 DLPC lipids. We adopt 25 peptides, namely melittin and Melp5, randomly and vertically inserted into the lipid bilayer as the respective initial structures (Fig.\ \ref{fig1}(c) and (d)). Each CG simulation has a duration of 5000 ns, during which the peptides undergo diffusion and thoroughly explore the entire simulation box in both the x and y directions.

Following the 5000 ns CG simulations, it is evident that the peptides exhibit a non-random distribution within the lipid bilayer. The observed oligomers exhibit sustained peptide organization conformation over a period of time, indicating stability or metastability. To acquire accurate statistical data pertaining to these oligomers, it is imperative to perform substantial sample measurements. Therefore, we conduct 20 replicated simulations of melittin and Melp5 respectively. These 20 replicas utilize identical parameters, except for varying random seeds. Each replica simulation iteration lasts 5000 ns. Spontaneous peptide aggregation is observed in both melittin and Melp5 simulations conducted for each replica. The number of peptides is 25, leading to a peptide-to-lipid (P/L) ratio of approximately $25/596\approx4.2/100$. The P/L ratio surpasses the minimum threshold value in previous studies \cite{54}, resulting in an accelerated rate of spontaneous aggregation in our simulations.

Through observing the MD trajectories to distinguish peptide aggregation in the membrane at final 100 ns, we find that the peptides can aggregate in different forms resulting in distinct oligomers, as depicted in Fig.\ \ref{fig2}. The statistical results pertaining to these oligomers are provided within brackets, and the name of each oligomer is shown below the images. The quantities of peptide aggregates differ considerably, with melittin having 14 oligomers and Melp5 having 28 oligomers. The varying levels of peptide aggregates indicate a higher propensity for peptide aggregate formation for Melp5 peptides.

In order to examine the factors contributing to the heightened peptide aggregation and enhanced pore formation effects observed in Melp5, our study focus on two specific oligomers. One oligomer is a tetramer of melittin, while the other oligomer is a pentamer of Melp5. In our work, they are designated as the NT- and PP-peptide, respectively. The NT- and PP-peptide have been selected due to their frequent occurrence in a significant proportion of our statistical results as oligomers. In this context, we provide a reason for excluding Melp5 trimers. The trimeric pore is very small, leading to limited water permeability. Despite the relatively highest possibility of trimer formation for Melp5 peptides, the defect of trimeric pores couldn't assist us to investigate the pore-forming mechanism. It is crucial to acknowledge that conducting a conformational analysis of the oligomer formation process aids in developing an intuitive understanding of peptide aggregation properties. These details will be discussed in subsequent subsections.

\begin{figure*}
\centering
\includegraphics[width=1.0\textwidth]{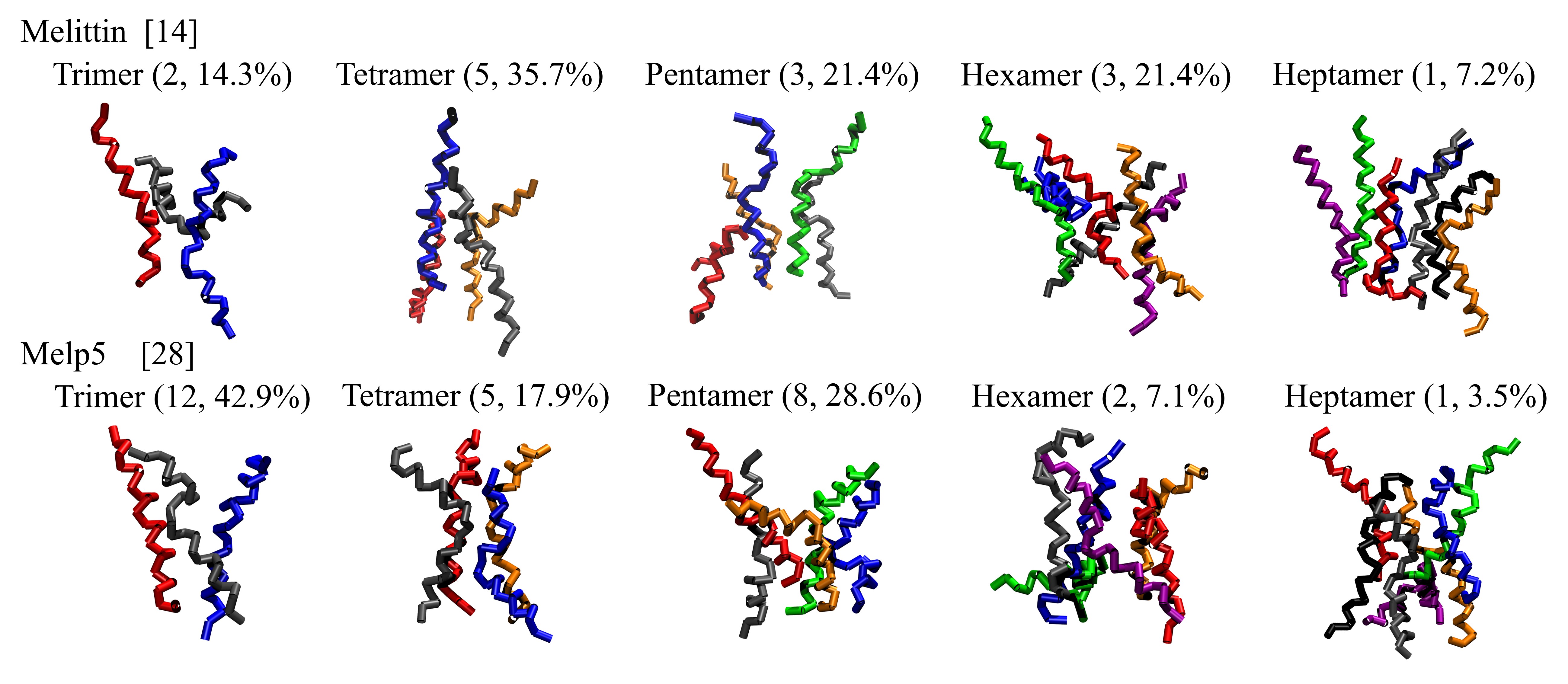}
\caption{
Multiple forms of oligomers are observed within the lipid bilayer subsequent to their spontaneous aggregation in 20 replicas of CG MD simulations. Each simulation comprises a total of 596 DLPC lipids and 25 peptides, which is conducted over a time period of 5000 ns simulations at the 310 K temperature. Each category of oligomer is accompanied by a bracket that displays the numerical values and corresponding percentages. The percentages represent the proportion of a specific oligomer among the oligomers in all statistical results.}
\label{fig2} 
\end{figure*}

\vspace*{0.5\baselineskip}
\textbf{Confirmation of the NT- and PP-peptides in CG simulations}
\vspace*{0.5\baselineskip}

Here, we examine the MD trajectories of conformational changes in the NT- and PP-peptide, as depicted in Fig.\ \ref{fig3}(a) and (b). The process of peptide aggregation necessitates the existence of a pre-oligomer and the incorporation of a monomer into the pre-oligomer system. The monomer, typically attached to the membrane surface, is inserted into the lipid bilayer through its hydrophobic N-terminal region. This integration process leads to the formation of a transmembrane state, wherein the monomer becomes attracted to the pre-oligomers and contributes to the formation of a new peptide aggregate system.

Fig.\ \ref{fig3}(a) and (b) illustrate the presence of pre-oligomers prior to the NT- and PP-peptide. The occurrences of the melittin trimer and Melp5 tetramer within the lipid bilayer are observed at 4910 ns and 4790 ns, respectively. The NT- and PP-peptide is formed through the introduction of a monomer. Fig.\ \ref{fig2} illustrates that Melp5 peptides have the capability to generate a substantial quantity of oligomers, especially trimers, which can serve as pre-oligomers for the formation of higher oligomers. This phenomenon enhances the possibility of higher oligomers being formed.

In order to gain a deeper understanding of why Melp5 peptides are able to aggregate to form more oligomers, we analyze the interpeptide and residue-membrane LJ interaction energy, as well as the residue-residue RDF. The subsequent sections will conduct these analyses.

\vspace*{0.5\baselineskip}
\textbf{Lennard-Jones potential stabilizes the oligomeric structure}
\vspace*{0.5\baselineskip}

Through our 5000 ns CG simulations, we have observed the spontaneous aggregation of peptides to form diverse oligomers. In order to gain insight into the distribution of these oligomers, it is imperative to investigate the fundamental factors that contribute to their occurrences. Under equilibrium conditions, the distribution of peptides in the membrane should follow the Boltzmann distribution, which implies that the oligomers possessing lower free energy will be more prevalent to form in membrane. Nevertheless, it is crucial to acknowledge that the formation of oligomers typically necessitates substantial free energy barriers. These barriers manifest as a result of interpeptide interactions, peptide to lipid interactions, as well as the conformational changes necessary for the process of aggregation. Therefore, both thermodynamic and kinetic effects are the dominant factors in the formation of different types of oligomers in the membrane.

In our work, we calculate the radial distribution function (RDF) within residues to residues. As depicted in Fig.\ \ref{fig4}(a) and (b), it can be observed that the highest peak of the $g(r)$ occur at about 0.5 nm, suggesting the interactions of residues to residues are dominated by LJ interactions. In addition, the curves follow a downward trend with increasing conformation complexity of oligomers. This result indicates that when the conformation of oligomers becomes more complex, the interactions of residues to residues are weakened, which could result in a progressive destabilization of the oligomeric structure. In addition, we notice that the peaks of $g(r)$ in the trimer and tetramer of Melp5 are 141 and 116 respectively, which are significantly higher than those (120 and 96) in melittin. In this way, we can infer that when the oligomeric structures are simpler, the residues are more tightly arranged, which indicates a greater pore stability and a greater possibility for peptides to spontaneously aggregate to form the oligomers.

\begin{figure*}[t!]
\centering
\includegraphics[width=0.8\textwidth]{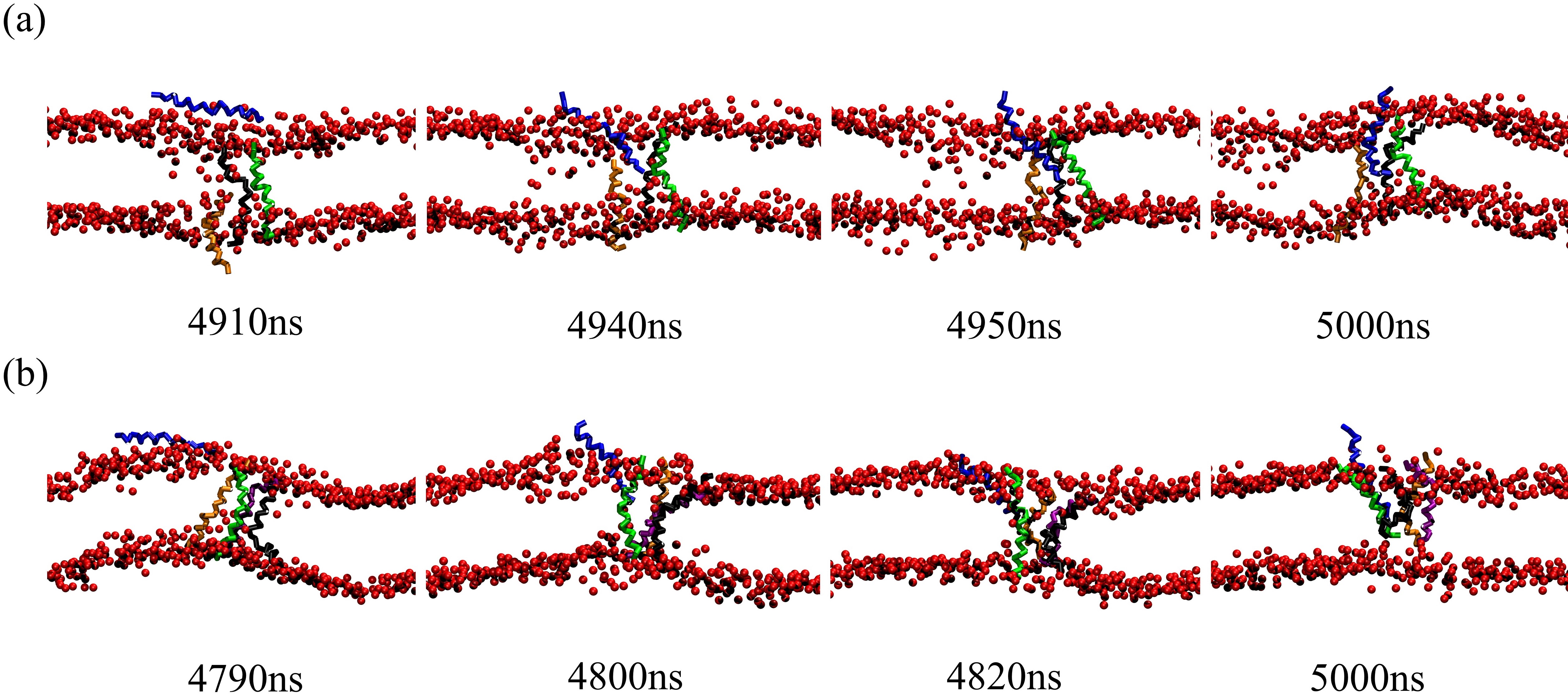}
\caption{
(a) The capture of snapshots during the process of a NT-peptide formation. The monomers of the NT-peptide exhibit distinct colors, namely blue, black, green, and orange. (b) The capture of simulation snapshots during a PP-peptide formation process. The monomers of the PP-peptide exhibit distinct colors, namely blue, black, green, orange, and purple. In order to enhance clarity, the lipid head groups are depicted as red balls while the lipid tail groups are omitted from the representation.}
\label{fig3} 
\end{figure*}

Interestingly, the statistical result shows that Melp5 pentamers have a higher probability of occurrence than Melp5 tetramers, where we use the interpeptide LJ interaction energy ccr}{Fig.\ \ref{fig4}(c) to investigate the potential reason for this result. As the oligomeric structure becomes more complex, the number of pairs of interpeptide interactions increases, which results in the values of interpeptide LJ interactions decreasing due to the superposition of calculations. It is worth noting that the difference in interpeptide LJ interaction energy between Melp5 tetramer and Melp5 pentamer is not obvious, with values of -498.34$\pm$157.67 kJ/mol and -500.62$\pm$156.14 kJ/mol respectively. This result may imply that from tetramer to pentamer of Melp5, the interpeptide barrier is easy to overcome, which could make the tetramer formation less thermodynamically favorable in comparison to the pentamer formation. Furthermore, Melp5 trimers have the highest probability of occurrence in the statistical results. It suggests that there is a considerable quantity of trimers as pre-oligomers that shows a distribution bias and the formation of tetramers is significantly limited due to the potentially substantial barriers to strong LJ interactions of residues to residues. Alternatively, the observed decrease in the tetramers could be attributed to their role as pre-oligomer states in the aggregation formation of pentamers.

In general, we compare $g(r)$ to investigate the reasons why Melp5 oligomers have a higher occurrence possibility compared to melittin, and discuss about Melp5 pentamers appear more frequently than tetramers. In order to further investigate the underlying factors for the enhanced poration ability and pore stability of Melp5, we analyze the effects of mutant residues, which are elaborated in the subsequent subsection.

\begin{figure*}[t!]
\includegraphics[width=0.7\textwidth]{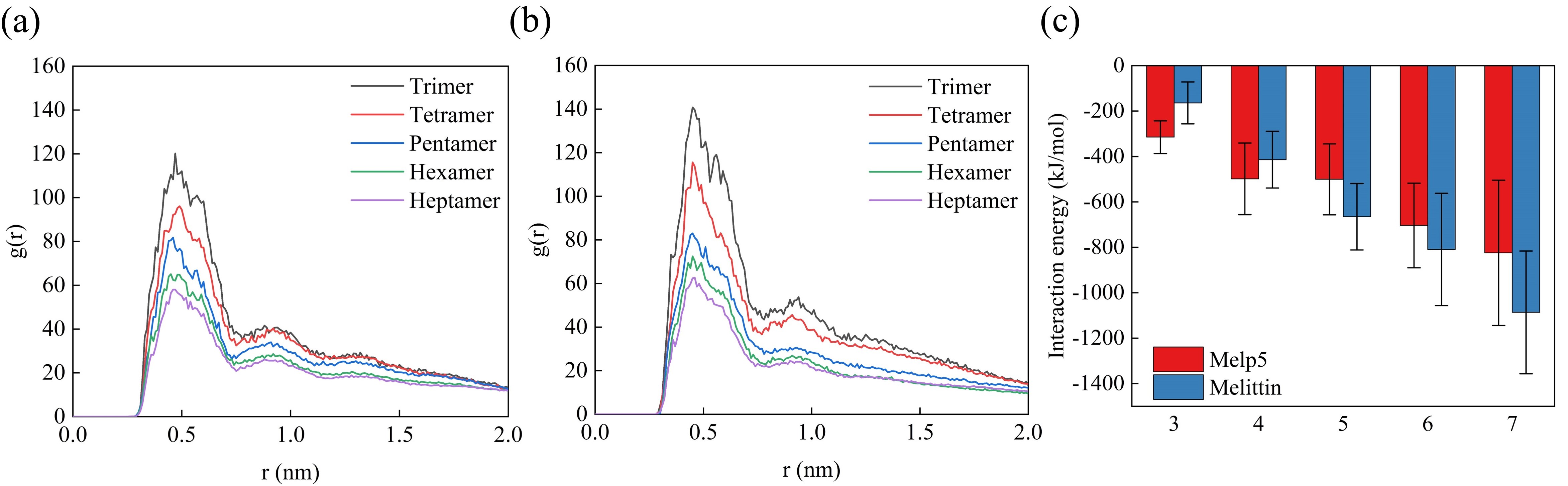}
\centering
\caption{
Radial distribution functions (RDFs) as $g(r)$ are plotted for melittin (a) and Melp5 (b) during the final 10 ns of CG simulations. RDFs of oligomers are differentiated by distinct color lines: the trimer is denoted by black, the tetramer by red, the pentamer by blue, the hexamer by green, and the heptamer by purple. (c) The interpeptide LJ interaction energy of distinct oligomers for melittin and Melp5.}
 \label{fig4} 
\end{figure*}

\vspace*{0.5\baselineskip}
\textbf{The effect of the mutant residues on pore stability}
\vspace*{0.5\baselineskip}

\begin{figure*}[t!]
\centering
\includegraphics[width=0.7\textwidth]{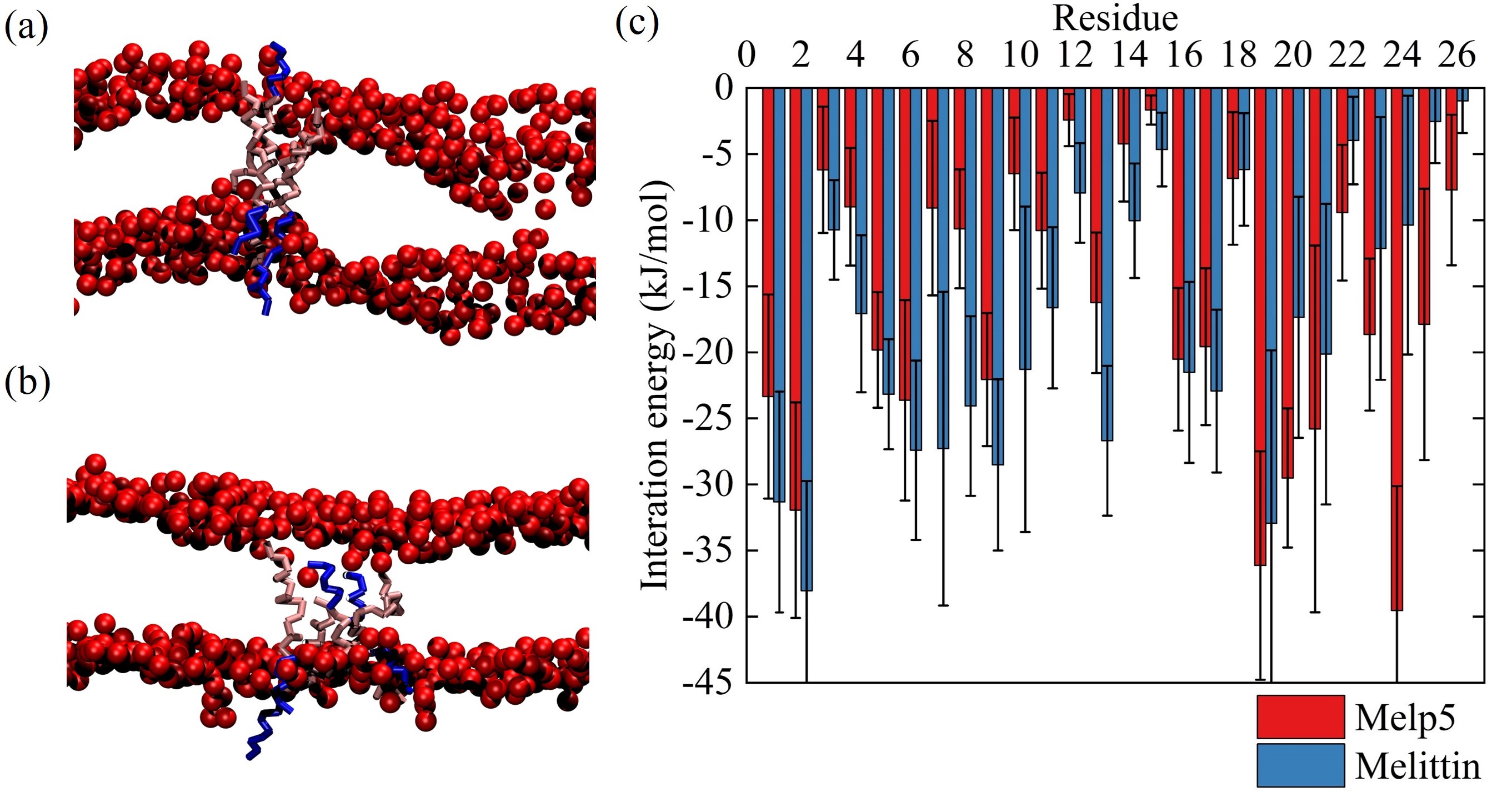}
\caption{The snapshots of NT- (a) and PP-peptide (b) at the final frame of CG simulations. The blue fragments correspond to the residues with the sequence range of 18 to 26. In order to enhance clarity, the lipid head groups are depicted as red balls while the lipid tail groups are omitted from the representation. (c) The distributions of LJ interaction energy between the residues and membrane. }
\label{fig5} 
\end{figure*}

\begin{table*}
\caption{The time of the PP-peptides with or without reversed mutant residues (ALA-22, GLN-24 and LEU-26) that maintain stable peptide organization conformation in the membrane for 5 replicated simulation results. The reversed mutant residues in amino acid sequence are highlighted in red.}
\centering
\setlength{\tabcolsep}{8.5mm}{
\begin{tabular}{lccccc} 
\toprule
\midrule
  \diagbox{Amino Acid Sequence}{Time}{Replicated Simulations} &   \multicolumn{1}{c}{1}  &   \multicolumn{1}{c}{2} &   \multicolumn{1}{c}{3} &   \multicolumn{1}{c}{4} &   \multicolumn{1}{c}{5} \\
  \midrule
  GIGAVLKVLATGLPALISWIKAAQQL &  210 & 195 & 90 & 87 & 51 \\
  GIGAVLKVLATGLPALISWIK\textcolor{red}{R}A\textcolor{red}{R}Q\textcolor{red}{Q} &  165 & 84 & 51 & 33 & 15  \\
  GIGAVLKVLATGLPALISWIK\textcolor{red}{R}A\textcolor{red}{R}QL &  84 & 72 & 30 & 21 & 15  \\
  GIGAVLKVLATGLPALISWIK\textcolor{red}{R}AQQ\textcolor{red}{Q} &  99 & 66 & 60 & 45 & 9  \\
  GIGAVLKVLATGLPALISWIKAA\textcolor{red}{R}Q\textcolor{red}{Q} &  201 & 45 & 36 & 30 & 6  \\
  GIGAVLKVLATGLPALISWIK\textcolor{red}{R}AQQL &  168 & 96 & 81 & 39 & 36  \\
  GIGAVLKVLATGLPALISWIKAA\textcolor{red}{R}QL &  141 & 117 & 60 & 42 & 39  \\
  GIGAVLKVLATGLPALISWIKAAQQ\textcolor{red}{Q} &  216 & 90 & 75 & 42 & 36  \\
  \midrule
  \bottomrule
  \end{tabular}}
  \label{tb2:table2}
\footnotesize{Times are in ns.} \\
\end{table*}

To further investigate the underlying factors that contribute to enhanced pore stability for Melp5 peptides, we focus on the interactions between residues and membrane. The MD trajectories used is 100 frames of the final 10 ns forming NT- and PP-peptide (Fig.\ \ref{fig5}(a) and (b)) oligomeric structure with sustained peptide organization conformation. From the observation of trajectories, the lipid head groups located at the inserted position of peptides exhibit noticeable collapse. In addition, it has been observed that the NT-peptide doesn't fully penetrate into the membrane, with some residues at the C-terminal remaining external to the membrane. In contrast, almost all residues of PP-peptide have penetrated the membrane. To assist us study the effects of mutant residues, we analyze the LJ interactions between the residues of a monomer and membrane (Fig.\ \ref{fig5}(c)). It is noteworthy to observe that, when considering the No.\ 18 residue as the threshold, the melittin residues in the front sequence (No.\ 1 to 17) exhibit higher interactions with the membrane. Conversely, the residues in the back sequence (No.\ 18 to 26), particularly the mutant residues (ALA-22, GLN-24 and LEU-26) of a Melp5 monomer, demonstrate significantly stronger interactions with the membrane. Mutant residues of the Melp5 monomer facilitate transmembrane penetration into the membrane, which could enhance the poration ability of Melp5 peptides through spontaneous aggregation.

We intend to reverse the mutant residues (ALA-22, GLN-24 and LEU-26) back into the original residues (ARG-22, ARG-24 and GLN-26) in a permutation or combination method to investigate how these mutations affect the oligomer-forming pores. The mutant residues of every peptide within PP-peptide are simultaneously reversed into the original residues as the amino acid sequences are shown in Table\ \ref{tb2:table2}. Here, the PP-peptides with or without reversed residues are reinserted into the lipid bilayer composed of 598 DLPC lipids by CHARMM-GUI online server. 0.1M NaCl is added to the system. We employ GROMACS to perform energy minimization in the NPT ensemble again. To reduce the contingency of simulation results, we conduct 5 simulation replicas for each amino acid sequence group. Each replicated simulation is conducted for a duration of 300 ns with the Martini 3.0 version force field, utilizing a time step of 2 fs and maintaining the temperature of 310 K.

The result shows that the PP-peptides without reversed residues (GIGAVLKVLATGLPALISWIKAAQQL) exhibit the most significant pore stability and maintain stable peptide organization conformation for more than 50 ns within 5 simulation replicas. The reversion of two or three specific mutant residues results in a notable decrease in the time of sustaining pore stability, where we could find the majority of the time below 50 ns. In addition, we can find that the time is even less than 10 ns in two cases (9 ns and 6 ns) when GLN-24 and LEU-26 or ALA-22 and LEU-26 are reversed. However, upon reversing only one residue (ALA-22 or GLN-24 or LEU-26), the majority of replicated simulation time don't exhibit significant differences in comparison to the time of the PP-peptides without reversion. But it is worth noting that when reversing only one residue, replicated simulation results with the time of less than 50 ns also could be found, which indicates it still affects pore stability with slight reduction. Therefore, our results indicate that pore stability is notably influenced by the exisitence of three specific mutant residues, namely ALA-22, GLN-24, and LEU-26. Nevertheless, in order to substantially improve the stability of the oligomer-formed pores, it is imperative to make three specific mutant residues existing in the Melp5 peptide at the same time.

\begin{figure*}[t!]
\centering
\includegraphics[width=0.8\textwidth]{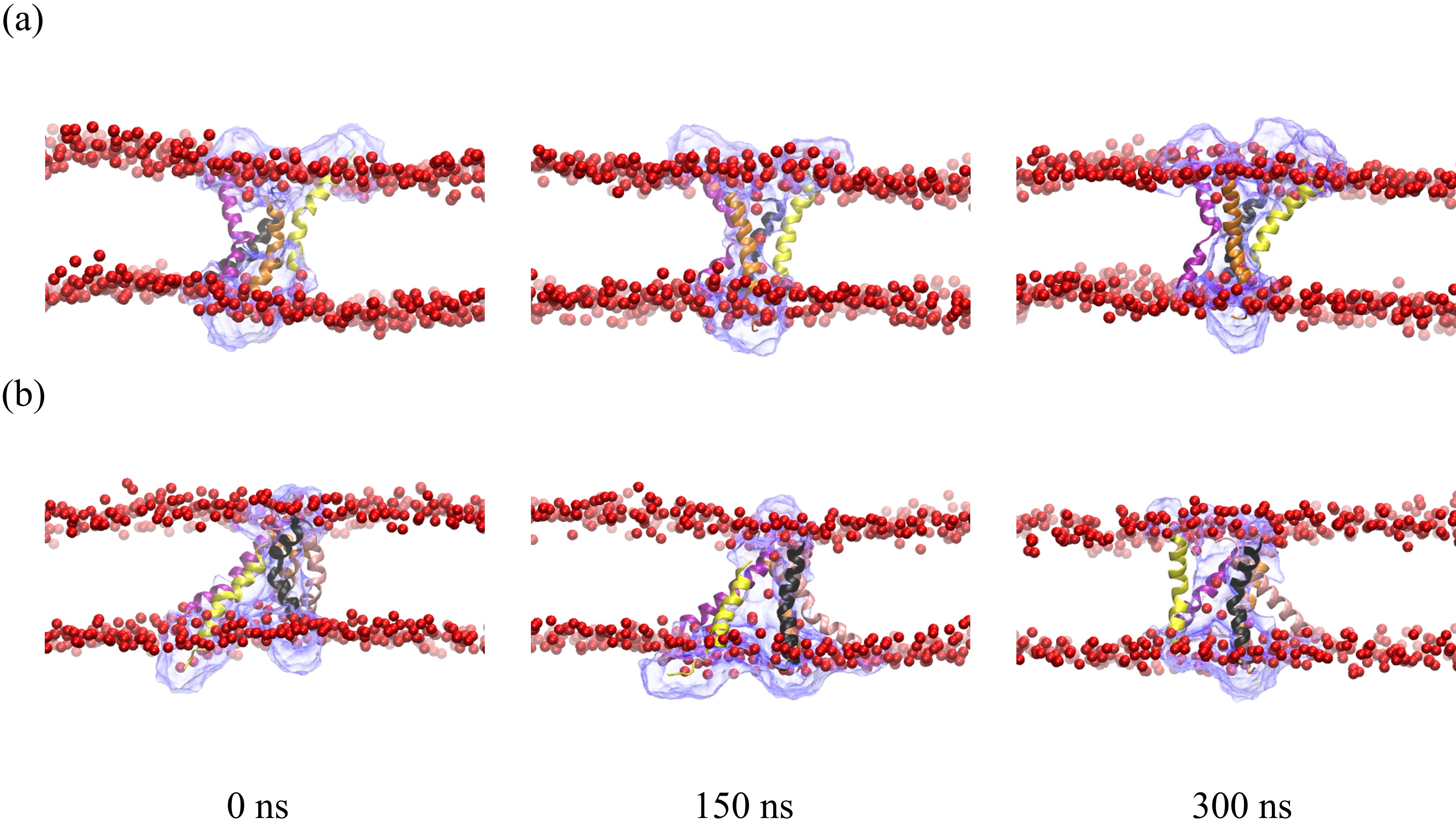}
\caption{All-atom simulation snapshots of the NT- (a) and PP-pore (b). The water channels are highlighted with blue surface. The lipid head groups are depicted as the red balls, while the water molecules located external to the lipid bilayer are represented by the green dashed lines. The lipid tail groups are omitted from the representation.}
\label{fig6} 
\end{figure*}

The omission of atomic structure and interaction at the atomic level in CG simulations necessitates the investigation of oligomer-formed pores using all-atom simulations. Previous research demonstrated that peptide aggregates can form water channels within the lipid bilayer. In the subsequent sections, we provide a description of the pores formed by the NT- and PP-peptide, referred to as the NT- and PP-pore, respectively, to obtain a comparison of the water permeability.

\vspace*{0.5\baselineskip}
\textbf{Confirmation of the NT- and PP-pores in all-atom simulations}
\vspace*{0.5\baselineskip}

The utilization of CG modeling in our simulations plays a significant role in capturing the phenomenon of spontaneous aggregation, which could help us to obtain relatively satisfactory results more quickly in limited computational resources. The Martini 3.0 version force field is employed in our CG simulations, as it is a widely accepted and standardized force field specifically designed for lipid systems. To verify the results of CG simulations, particularly regarding the stability of the NT- and PP-pores, we also conduct all-atom simulations. The dimensions of the simulation box in our work are estimated to be approximately 14 × 14 × 8 nm. The NT- and PP-peptide's final structures obtained from CG simulations are chosen and converted back to atomistic structures using the backmapping software in CHARMM-GUI. Here, we use the CHARMM36M force field to conduct all-atom simulations that incorporate explicit water molecules.

\begin{table}[b!]
\caption{The electrostatic interaction energies between C-terminal residues and DPPC/POPG lipids are investigated using all-atom simulations.}
\centering
\setlength{\tabcolsep}{1.5mm}{
\begin{tabular}{llll} 
\toprule
\midrule
  \diagbox{Residue}{Electrostatic}{Peptide} &   \multicolumn{1}{c}{Melittin}  &   \multicolumn{1}{c}{Melp5} \\
  \midrule
  21st &  -167.94$\pm$91.46 & -64.00$\pm$52.01  \\
  22nd &  -118.26$\pm$83.10 & \ 0.41$\pm$3.16  \\
  23rd &  -72.64$\pm$92.43 & -1.25$\pm$8.38  \\
  24th &  -201.81$\pm$75.39 & -15.72$\pm$15.88  \\
  25th &  -20.52$\pm$19.30 & -13.49$\pm$14.88  \\
  26th &  -10.07$\pm$17.35 & -12.32$\pm$19.25  \\
  \midrule
  \bottomrule
  \end{tabular}}
  \label{tb3:table3}
\footnotesize{Electrostatic interaction energies are in kJ/mol.} \\
\end{table}

\begin{figure*}
\centering
\includegraphics[width=0.7\textwidth]{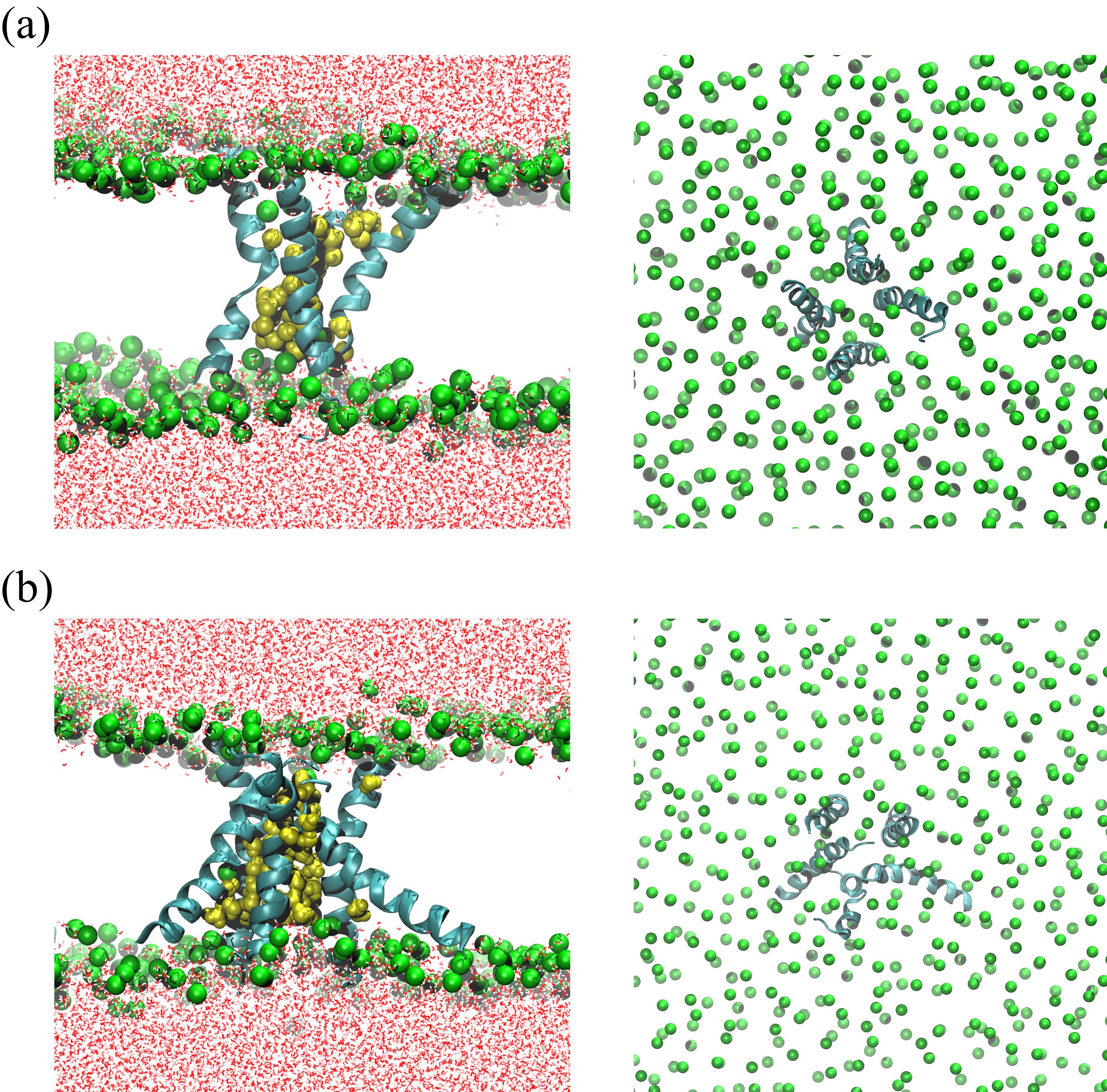}
\caption{The water molecules inside the NT- (a) and PP-pore (b) at the final frame of 300 ns all-atom simulations. The lipid head groups are depicted as green balls, while the water molecules within the pores are visually highlighted with yellow balls. The lipid tail groups are omitted from the representation. NT- and PP-pore's top view of snapshots are shown in the right.}
\label{fig8} 
\end{figure*}

\begin{figure*}[t!]
\centering
\includegraphics[width=0.7\textwidth]{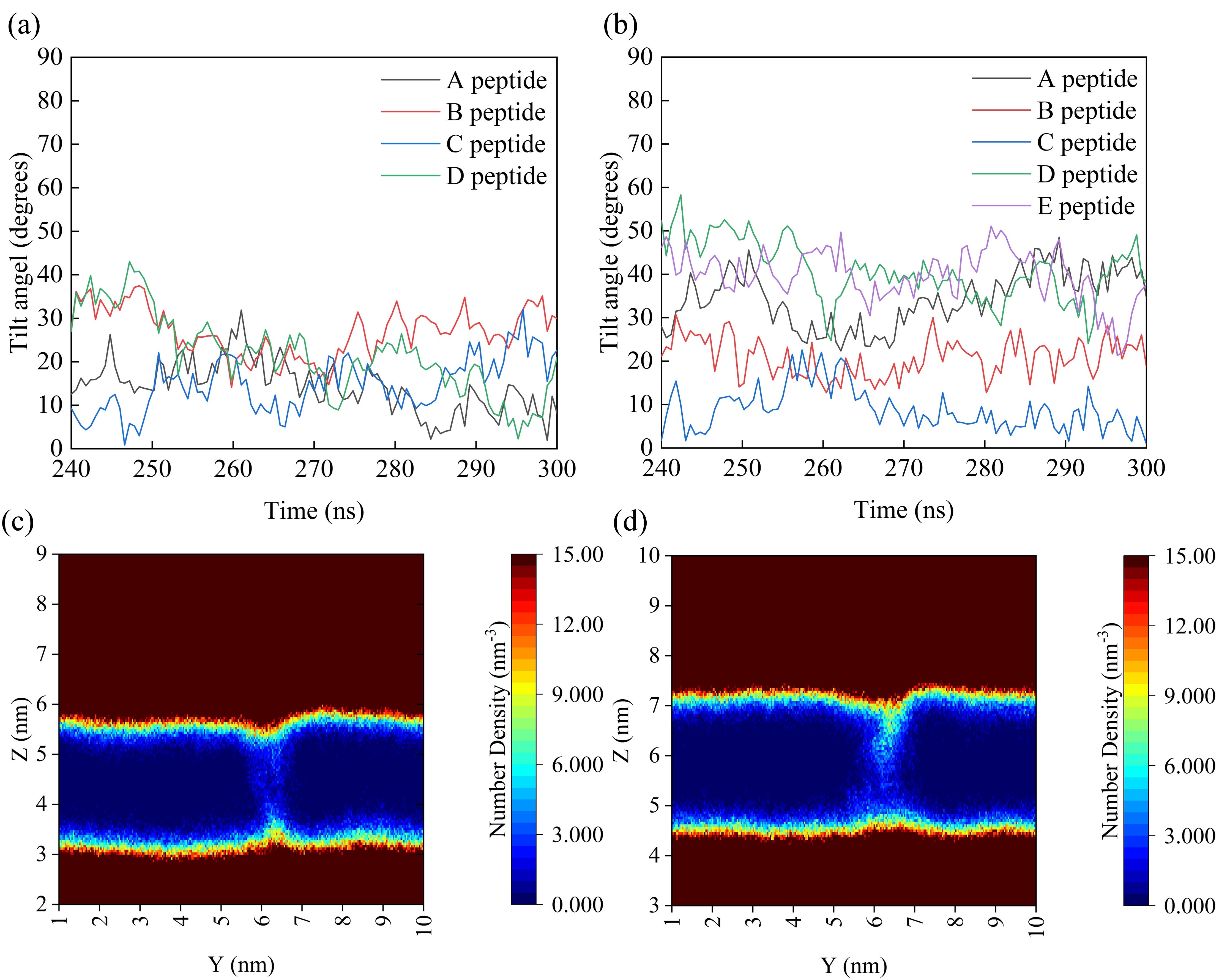}
\caption{The fluctuation curves of tilt angle for each monomer of the NT- (a) and PP-pore (b) at final 60 ns in all-atom simulations. The fluctuation curves are represented by different colored lines, which are represented by black, red, blue and green lines in (a) and represented by black, red, blue, green and purple lines in (b). Two dimensional water density maps plot in the y-z plane for the NT- (c) and PP-pore (d).}
\label{fig7} 
\end{figure*}

During the 300 ns all-atom simulations, both the NT- and PP-pore can remain stable conformations throughout, which are captured as snapshots depicted in Fig.\ \ref{fig6}(a) and (b), which confirms the pore stability. The water channels demonstrate structurally stable configurations, wherein the hydrophilic residues are oriented inwardly towards the aqueous pore, while the hydrophobic residues are directed outwardly towards the lipid bilayer. The monomers of oligomers are not densely packed, which allows for the presence of negatively charged phospholipid head groups to bend into the inner lining of the pore, resulting in a peptide-membrane system with enhanced stability.

The peptide helicity and hydrogen bonds during all-atom simulations can be analyzed by GROMACS. It is noteworthy to mention that hydrogen bonds and helicity significantly influences the pore stability. The helicity values obtained for the NT-pore and PP-pore are 77.7$\pm$1.70\% and 81.3$\pm$0.94\%, respectively. The NT- and PP-pore exhibit hydrogen bonds counts of 66$\pm$1 and 84$\pm$1, respectively. The results provide evidence that PP-pore has a higher helicity value which predominantly adopts a more $\alpha$-helix structure than NT-pore. The helicity of the NT-pore is comparatively lower than that of the PP-pore, potentially due to the increased occurrence of hydrogen bonding of residues to residues within the PP-pore caused by mutant residues. This increased hydrogen bonding leads to a more stable $\alpha$-helix secondary structure, thereby enhancing helicity.

\vspace*{0.5\baselineskip}
\textbf{Water permeability and pore radius of the NT- and PP-pore in all-atom simulations}
\vspace*{0.5\baselineskip}

The antimicrobial effectiveness of AMPs is heavily dependent on the water permeability within the pores. Water permeability can be analysed by capturing water molecules in the all-atom simulations. We observe a substantial quantity of water molecules in both the NT-pore and PP-pore, as depicted in Fig.\ \ref{fig7}(a) and (b). The average quantities of water molecules, denoted as $N_{w}$, in the NT- and PP-pore are observed as 76$\pm$23 and 57$\pm$13, respectively. By utilizing the approximation $H\approx$ 2 nm and substituting $N_{w}$ into the equation, we obtain the values of $r_{PP-pore}=$ 1.20$\pm$0.66 nm and $r_{NT-pore}=$ 1.04$\pm$0.49 nm for the NT- and PP-pore, respectively. This result suggests that the PP-pore exhibits enhanced water permeability with a larger radius of the pore compared to the NT-pore. 

Furthermore, we notice that the monomers of a PP-pore have larger tilt angles compared with those of a NT-pore. Thus, we use the final 60 ns where the pores could keep stable positions in the membrane to conduct an analysis of the tilt angle for each monomer of the NT- and PP-pore (Fig.\ \ref{fig8}(a) and (b)). The tilt angles of all monomers in the NT-pore fluctuation always below 40$^\circ$, with average values of 26.9$^\circ$ and 21.7$^\circ$ for the B-peptide and D-peptide, respectively. Nevertheless, the D-peptide and E-peptide in the PP-pore show notably larger average tilt angles, which are 41.0$^\circ$ and 39.8$^\circ$, respectively. In order to visually observe the configurations of water channels in the NT- and PP-pore, we generate water density maps in the yz plane, as depicted in Fig.\ \ref{fig8}(c) and (d).

It should be noted that all-atom simulations accurately capture electrostatic interactions more accurately than CG simulations. Despite the inherent limitations of accurately representing electrostatic interactions in CG simulations, spontaneous aggregation and pore formation still could be manifested. Subsequently, the pores could be observed to exhibit the stability in all-atom simulations. In general, the examination of pore formation within our CG simulations and the verification of pore stability through our all-atom simulations can gain valuable insights into the influence of electrostatic interactions on poration processes.

We calculate electrostatic interaction energy between the C-terminal residues and the DPPC/POPG lipids (Table\ \ref{tb3:table3}). It is known that phospholipids are negatively charged, which can specifically prompt the positively charged residues (LYS-21, ARG-22, LYS-23, and ARG-24) of melittin to exhibit enhanced electrostatic interactions with phospholipids. Due to the positive charge present in the N-terminal residues, the C-terminal and N-terminal residues of a melittin monomer adhere to the lipid bilayer surface at a slight tilt angle. The C-terminal mutant residues (ALA-22, ALA-23 and GLN-24) of a Melp5 monomer have lost their positive charge. As a result, they exhibit reduced affinity for the head groups of phospholipids present on the lipid bilayer surface, making their adsorption less favorable. Therefore, the C-terminal of a monomer exhibits more unrestricted movement and rotation within the lipid bilayer, leading to a comparatively significant tilt angle. Fig.\ \ref{fig9}(a) and (b) depict the simulation snapshots of a melittin monomer and a Melp5 monomer, respectively, in the final frame.

\begin{figure}[t!]
\centering
\includegraphics[width=0.48\textwidth]{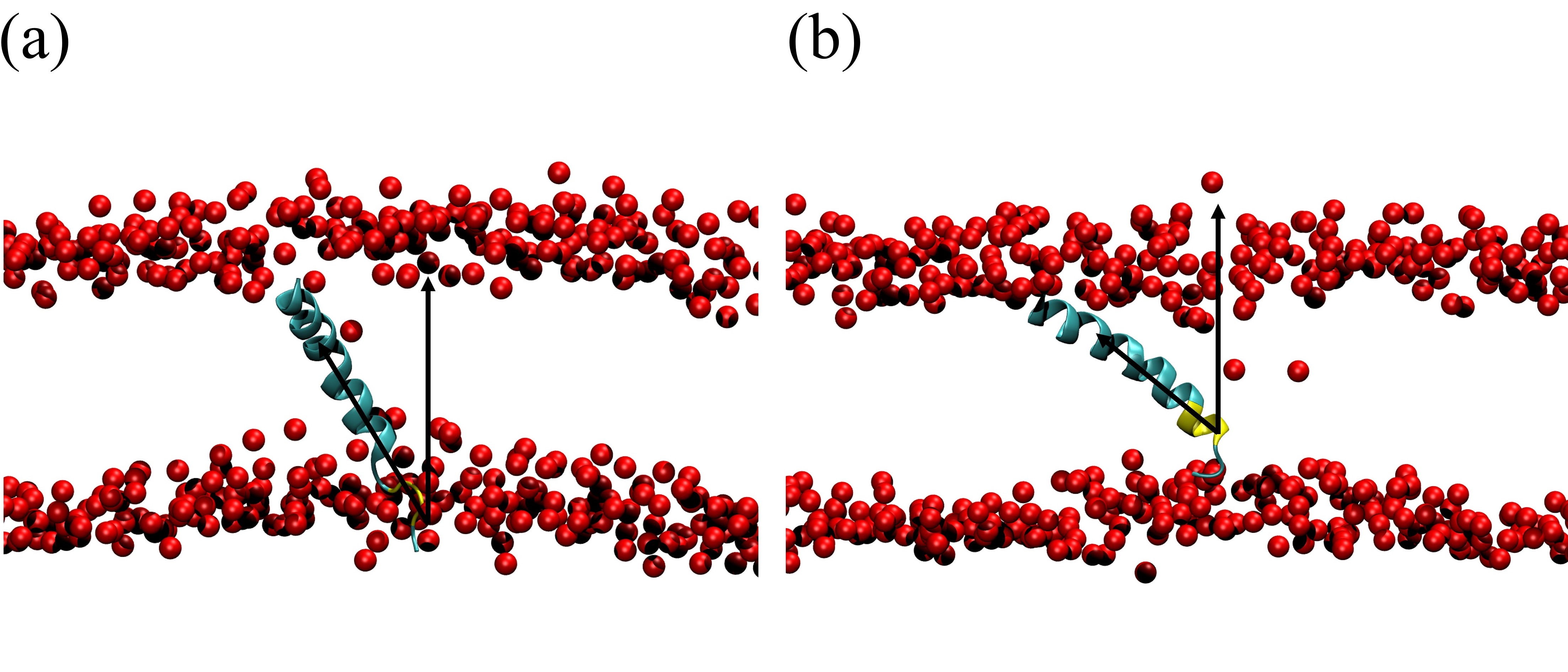}
\caption{The final frame captured snapshots of a melittin monomer (a) and a Melp5 monomer (b) after 300 ns all-atom simulations. The lipid head groups are depicted as red balls, residues with sequential numbers 22 to 24 are highlighted in yellow. The lipid tail groups are omitted from the representation.}
\label{fig9} 
\end{figure}

\vspace*{0.5\baselineskip}
\textbf{Comparison with previous studies}
\vspace*{0.5\baselineskip}

Previous studies have examined CG or all-atom MD simulations to investigate the phenomenon of spontaneous aggregation of melittin and Melp5 within the lipid bilayer, which results in the formation of transmembrane pores. However, our work exhibits notable distinctions from previous studies in various aspects. First, in the work conducted by Yuan et al.\ \cite{30}, the AMBER ff99SBildn force field was employed to investigate the melittin aggregates within a mixed bilayer composed of DOPC/DOPG for the all-atom simulations. The all-atom molecular MD simulations in our work employ the CHARMM36M force field, while the lipid layer consists of the mixed DPPC/POPG lipid bilayer, as utilized in our previous work \cite{11}. Second, Santo et al.\ \cite{18}. carried on CG MD simulations by placing the melittin peptides attached to a single leaflet of the membrane surface as initial conformation. However, in our work, both melittin and Melp5 peptides are inserted perpendicularly into the lipid bilayer, which results in accelerating the formation of oligomers in membranes. Third, Yang et al.\ \cite{55} conducted a comparison to determine whether melittin and its variants exhibited aggregation and formed stable transmembrane pores at different P/L ratios. Our work is to perform statistical evaluation on the peptide aggregates generated by variant Melp5 and melittin at a specific P/L ratio in order to investigate the distinction in poration ability. We conduct a comparative analysis of our results with those presented in previous studies. The distribution of LJ interaction energy between the residues and the lipid bilayer in our work exhibits similarities to the results described by Yang et al.\ \cite{55}. In addition, our simulation results are relatively consistent with previous studies \cite{17,30,38} when examining the pore radius and water permeability of NT- and PP-pores. And the investigation of monomer tilt angle in our work refers to the previous studies conducted by Pino-Angeles et al.\ \cite{24,25}, where the range of fluctuation in the tilt angle is comparatively similar. In general, our analysis of the simulation results demonstrates a comparative consistency with previous study results. It is noted that our work focuses on investigating residue-residue and residue-membrane interactions that could be correlated with the conformation of oligomers. We aim to emphasize the influence of these interactions on pore stability and poration ability. Since Melp5 peptides have been observed to have an enhanced ability for pore formation, it is imperative to undertake a study that specifically focuses on the mutant residues, which can help us understand the underlying reasons for the enhancement in antimicrobial ability.

\section{Conclusion}

In our work, we have observed the distinct oligomers formed by melittin and Melp5 in lipid bilayer following the spontaneous aggregation during 5000 ns CG simulations. It is obviously apparent that Melp5 exhibits an increased propensity for higher oligomers compared to melittin, particularly with regards to forming the pentamers, while melittin is likely to form the tetramers. Our attention is directed towards the Melp5 pentamer (PP-peptide) and melittin tetramer (NT-peptide) for analysis. Every category of oligomer can create a transmembrane pore in the peptide-membrane environment. Two distinct pores, namely the NT- and PP-pore, which are the aqueous pores formed by the NT- and PP-peptide, have been chosen for meticulous investigation and analysis in all-atom simulations.

To investigate the enhancement of pore ability for Melp5, we employ the RDFs as $g(r)$ to analyze the reasons contributing to the increased possibility of oligomer formation. Furthermore, combined with the interpeptide LJ interactions, we discuss the potential factors about why Melp5 peptides are inclined to form pentameric structures instead of tetrameric structures. Furthermore, we focus on the mutant residues effect of pore stability by calculating the residue-membrane LJ interaction energy. Our results indicate that the mutant residues (ALA-22, GLN-24 and LEU-26) of Melp5 peptides show remarkably strong LJ interactions with the membrane, which could contribute to the Melp5 oligomeric enhanced pore stability. 

The confirmations of the NT- and PP-pore have been observed through 300 ns all-atom MD simulations. There are more hydrogen bonds within PP-pore, which results in enhanced stability and helicity in its secondary structure. In addition, the water permeability of the PP-pore is enhanced due to its larger pore radius. Moreover, we calculate the residue-membrane electrostatic interaction energies which shows that the mutant residues (ALA-22, ALA-23 and GLN-24) could exhibit significantly diminished electrostatic interactions with membrane in comparison to the original residues (ARG-22, LYS-23 and ARG-24). The difference in electrostatic interactions could cause Melp5 monomer to have a more substantial tilt angle in membrane.

Both melittin and Melp5 can undergo spontaneous aggregation, resulting in the formation of oligomers. Melp5 has been shown to facilitate the formation of enlarged pores, thereby enhancing water permeability. However, the mechanism for Melp5 peptides spontaneous aggregation into higher oligomers with enhanced poration ability are not sufficiently understood and explained. Our work is aimed at gaining a deeper comprehension of the critical role of mutant residues for the formation of large transmembrane pores and enhancement of pore stability to improve antimicrobial ability. Therefore, in the future, the antimicrobial ability of AMPs can be effectively enhanced by designing mutant residues that promote peptide aggregation and stabilization of transmembrane pores.


\bibliography{reference}

\end{document}